\documentclass[useAMS,usenatbib]{mn2e}
\usepackage{graphicx,rotate,url,mathptmx,lscape,times,amssymb,color}
\def\gtsim{\mathrel{\hbox{\rlap{\hbox{\lower4pt\hbox{$\sim$}}}\hbox{$>$}}}}
\def\lesssim{\mathrel{\hbox{\rlap{\hbox{\lower4pt\hbox{$\sim$}}}\hbox{$<$}}}}

\def\Msun{M$_{\odot}$}
\def\cm{{\rm\thinspace cm}}

\def\ga{{\rm\thinspace gauss}}

\def\Msun{\hbox{$\rm\thinspace M_{\odot}$}}

\def\s{{\rm\thinspace s}}

\def\ha{\hbox{{\rm H}$\alpha$}}
\def\hb{\hbox{{\rm H}$\beta$}}

\def\oiii{\hbox{{\rm [OIII]\ }}}

\def\nii{\hbox{{\rm [NII]\ }}}
\def\h0{\hbox{{\rm H}$^0$}}
\DeclareMathAlphabet{\vib}{OML}{cmm}{m}{it}

\title[Broad and Luminous \oiii and \nii in GC ULXs]{Broad and Luminous \oiii and \nii in Globular Cluster ULXs}

\author[R.L. Porter et al.]
       {\parbox[]{6.0in}
        {R.L. Porter\thanks{E-mail: rlporter@umich.edu}\\
        \footnotesize
        Department of Astronomy, University of Michigan, 500 Church Street, Ann Arbor, MI, 48109-1042, USA\\}}
\date{
      Received }

\begin{document}

\maketitle

\label{firstpage}

\begin{abstract}
We consider an accretion-disc origin for the broad and luminous forbidden-line emission observed in 
ultraluminous X-ray (ULX) sources CXOJ033831.8-352604 and XMMU 122939.7+075333 in globular clusters hosted
by elliptical galaxies NGC~1399 and NGC~4472, respectively.
We will refer to the latter by the globular cluster name RZ2109.  
The first has strong [OIII] and [NII], the second only [OIII].
Both H$\alpha$ and H$\beta$ are very weak or undetected in both objects.
We assume that the large line widths are due to Keplerian rotation around a compact object
and derive expressions for maximum line luminosities.
These idealized models require central masses $\gtrsim100$ and $\gtrsim30000\Msun$ for CXOJ033831.8-352604 and RZ2109, respectively.
An independent, bootstrap argument for the total disc mass yields, for both systems,
$M_{\mathrm{disc}}\gtrsim10^{-4}\Msun$ for a purely metallic disc
(and two orders of magnitude larger for solar metallicities).
If Roche-lobe overflow is implicated, viscous time-scales are $\gtrsim300$~yr.
Standard disc theory then offers another limit on the central masses.
Lobe radii for a $\sim1\Msun$ donor are $\gtrsim10^{13}$~cm.
We therefore rule out Roche-lobe overflow of a white dwarf in both systems.
Red giants could fill the necessary lobes.  
Whether they are too metal-poor to produce the strong forbidden lines without strong hydrogen emission is unclear.
\end{abstract}

\begin{keywords}
 galaxies: individual (NGC 1399) --  galaxies: individual (NGC 4472) -- galaxies: star clusters -- accretion, accretion discs.
\end{keywords}

\section{Introduction}

Ultraluminous X-ray sources are point sources that appear to exceed the Eddington luminosities
of stellar-mass compact objects.  They are therefore prime candidates for long-sought
intermediate-mass black holes (IMBHs).  ULXs have been studied extensively since their
discovery in the early 1980s, and they continue to be the subjects of intense research activity
(Miller, Fabian, \& Miller 2004; Winter, Mushotzky, \& Reynolds 2006; 
Hui \& Krolik 2008; King 2008, 2009).

Zepf et al. (2007; 2008, hereafter Z08) and Irwin et al. (2010, hereafter I10) recently identified 
very broad and luminous \oiii $\lambda5007$ emission in two ULXs harbored by globular clusters
in elliptical galaxies NGC 4472 and NGC 1399, respectively.
The connections to globular clusters are robust, as in both I10 and Z08,
we see large redshifts that agree very well with the redshifts of the clusters and their host galaxies.  
Line widths were 140 and 1500 km~s$^{-1}$ in the two systems, and line luminosities, 
assuming the sources radiate isotropically, were $10^{36-37}$~erg~s$^{-1}$.
Both Z08 and I10 found difficulty simultaneously explaining the line widths and luminosities
with an accretion disc around a stellar black hole, 
the most natural explanation for a consistent source
with $L_x \approx10^{39}$~erg~s$^{-1}$.
The observed lines created a tension.  
Their large line widths suggested rapid rotation very close to the source,
while the large luminosities suggested a large line-emitting region and an origin in the outer disc.
The lack of \ha\ and \hb\ presented further complications in the context of globular clusters.
Z08 suggested a stellar black-hole wind is the explanation for RZ2109.
I10 suggested for CXOJ033831.8-352604 tidal disruption of a white dwarf by an IMBH
(Fabian, Pringle, \& Rees 1975; Rosswog, Ramirez-Ruiz, \& Hix 2009).

This Letter is organized as follows.
In section~\ref{maxlum}, 
we revisit and extend the arguments presented by Z08 and I10,
first relaxing the assumption that the line-emitting gas cannot be more dense than the critical density of the observed lines.
We derive a simple expression for the maximum line luminosity in an accretion-disc geometry.
Our expression is independent of temperature, density, and metallicity,
and depends only on rotational velocity and central mass.
While a super-critical density alleviates some of the tension described above, 
we nonetheless confirm the Z08 and I10 conclusions that stellar-mass accretion cannot be implicated in either ULX. 
In section~\ref{discmass}, we present bootstrap arguments for the total mass in an accretion disc and obtain lower limits.
In section~\ref{roche}, we explore the implications for Roche-lobe overflow.

\begin{figure}
\protect\resizebox{\columnwidth}{!}
{\includegraphics[angle=0]{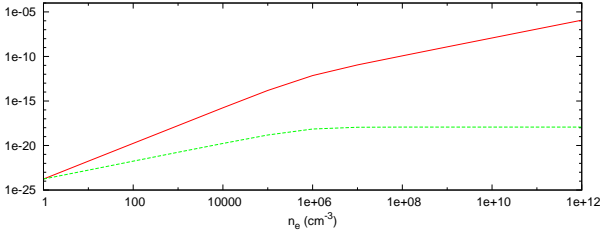}}
\caption{\oiii $\lambda5007$ emissivity (red solid, erg~cm$^{-3}$~s$^{-1}$) and intensity (green dashed, erg~cm$^{-2}$~s$^{-1}$) 
emitted in an isothermal, constant-density line-of-sight with solar metallicity and column density $1~\rm cm^{-2}$.  
Each curve is a function of volume density.
The critical density can be seen as a ``knee'' in each curve near $n_e = 10^{6}~\rm cm^{-3}$.}
\label{fig:emis}
\end{figure}

\section{Maximum Line Luminosities}
\label{maxlum}

Here we derive expressions for the maximum line luminosities of the \oiii and \nii forbidden lines.
We focus on the \oiii lines because they are seen in both objects.
Because both \nii and \oiii are produced by ions of the carbon isoelectronic sequence,
our formalism applies to both.
The final expression is independent of temperature, density, and metallicity and 
depends only on rotational velocity and central mass. 

We assume the gas is photoionized and collisionally excited and
begin by writing a simple two-level balance equation 

\begin{equation}
n_e n_1 q_{12} = n_e n_2 q_{21} + n_2 A_{21},
\label{balance}
\end{equation}

\noindent where $n_1$ and $n_2$ are the populations of the lower and upper levels, 
$A_{21}$ is the spontaneous radiative transition probability ($\s^{-1}\,$) from level 2 to level 1, 
and $q_{12}$ and $q_{21}$ are the collisional excitation and de-excitation coefficients ($\cm^{3}\s^{-1}\,$) 
and are related by $q_{12} = \frac{\omega_2}{\omega_1} q_{21} \exp(-h\nu/kT)$.  
The critical density $n_{\mathrm{crit}} \equiv A_{21}/q_{21}$ is the density at which the collisional de-excitation rate ($n_e q_{21}$) 
equals the radiative decay rate.
For $n_e \ll n_{\mathrm{crit}}$, the radiative decay dominates and the population of level 2 is

\begin{equation}
n_2 = \frac{n_e n_1 q_{12}}{A_{21}}.
\label{pop2}
\end{equation}
\noindent For $n_e \gg n_{\mathrm{crit}}$, 
the usual Boltzmann equation of local thermodynamic equilibrium (LTE) applies.
The local emissivity is $\epsilon = n_2 h\nu_{21} A_{21}$, and it follows that 
\begin{equation}
\epsilon = \Bigg\{ \begin{array}{ll} 
n_e n_1 q_{12} h \nu_{21} & \mbox{$n_e \ll n_{\mathrm{crit}}$} \\ 
\\
n_1 \frac{\omega_2}{\omega_1} \exp(-h\nu/kT) h\nu_{21} A_{21} & \mbox{$n_e \gg n_{\mathrm{crit}}$} .
\end{array} 
\Bigg.
\label{emiss} 
\end{equation} 
\noindent The critical density marks the transition from the regime where radiative cooling is proportional to $n_e~n_{\mathrm{ion}}$ 
to the LTE regime where cooling is proportional to $n_{\mathrm{ion}}$.  
The emitted energy per unit volume per unit time increases monotonically with increasing density. 
The two regimes are illustrated in Figure~\ref{fig:emis} for the \oiii $\lambda5007$ line.  
The critical density is represented by the knee near $10^{6}$~cm$^{-3}$.

Both I10 and Z08 considered the low-density case, and we have no need to reconsider that here.
The remainder of this work will assume $n_e > n_{\mathrm{crit}}$.
Authors who have discussed the high-density behavior of \oiii emission include 
Nussbaumer \& Storey (1981), Keenan \& Aggarwal (1987), 
Kastner \& Bhatia (1989), 
and Osterbrock \& Ferland (2006).  
From an observational perspective, we note, for example, Andre\"{a}, Dreschel, \& Starrfield (1994) 
found $n_e$ as high as $10^{8}$cm$^{-3}$ from \oiii lines in classical novae.

We generalize the high-density emissivity to multi-level configurations by using
the O$^{+2}$ calculations of Nussbaumer \& Storey (1981).
Their calculations are valid for all practical densities and include temperatures as high $40,000$~K,
safely above the temperature of photoionized O$^{+2}$ gas considered in section~\ref{discmass} below.
The upper level of the \oiii $\lambda5007$ and \nii $\lambda6548$ transitions is $^1D_2$.
The fractional population $f(^1D_2)$ in Table 5 of Nussbaumer \& Storey does not exceed $\approx0.2$.
A comparable value applies to N$^+$, and our emissivities take the maximum value
\begin{equation}
\epsilon_{max} =  n_A f(^1D_2)_{max} h\nu_{21} A_{21},
\label{final_emis_max}
\end{equation} 
\noindent where $n_A$ is the density of the ionization stage.

Next we posit a column of gas having local emissivity given by Equation~\ref{final_emis_max}.  
The intensity emitted through the column is
\begin{equation}
I_{\lambda} = \int \epsilon dz,
\label{eqn:intensity}
\end{equation}
\noindent where $z$ represents position along the column 
(or above an annulus perpendicular to the plane of the accretion disc).
To impose an upper limit on the integration variable,
we introduce a photon escape probability, $P_{\mathrm{esc}} = \frac{1}{1+\tau}$, 
where $\tau$ is the (Napier) line-center optical depth.
Equation~\ref{eqn:intensity} is then written
\begin{equation}
I^{max}_{\lambda} = N_A f(^1D_2)_{max} h\nu_{21} A_{21} P_{\mathrm{esc}},
\label{eqn:maxintensity} 
\end{equation}
\noindent 
where we have used $N_{A} = n_A \int dz$.  

\begin{figure}
\protect\resizebox{\columnwidth}{!}
{\includegraphics[angle=0]{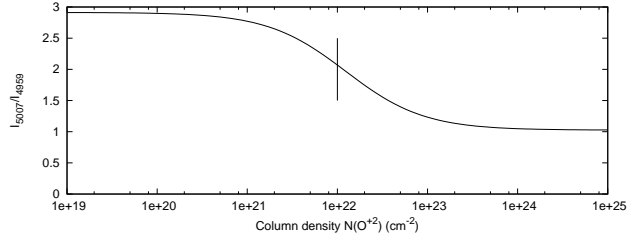}}
\caption{Theoretical ratio of the lines in the \oiii doublet $\lambda \lambda 4959,5007$ as a function of O$^{+2}$ column density.  
Observations in both Z08 and I10 clearly suggest ratios toward the canonical optically-thin value.   
The vertical bar marks $N$(O$^{+2}$)$ = 10^{22}$~cm$^{-2}$ (or $\tau_{5007} = 1$).}
\label{fig:doublet}
\end{figure}

Equation~\ref{eqn:maxintensity} saturates at $\tau \approx 1$,
consistent with the rule-of-thumb that we can see into a cloud only up to optical depth unity  (e.g., Rybicki \& Lightman 1979).
The upper limit corresponds to $N($O$^{+2}$) $\approx 10^{22}$~cm$^{-2}$, assuming local line widths are thermal at $\approx10^{4}$~K.
Optical depth in the \oiii lines causes their relative strengths to decrease from the canonical 3:1 ratio to approximate parity in the optically-thick limit.
This behavior is shown in Figure~\ref{fig:doublet}.  The plotted ratio is visibly greater than unity in both Z08 and I10.
Our limit, therefore, has the consequence of preventing model line-ratios inconsistent with the observations.   

Note that Equation~\ref{eqn:maxintensity} does not depend upon density.  
The green dashed curve in Figure~\ref{fig:emis} illustrates this and is an important point in our analysis.  
Above the critical density, the intensity emitted through a fixed column density is independent of the volume density of the gas.
This means that, with no penalty on the total intensity, we can increase the density and, with all else fixed,
squeeze the gas into a smaller volume,
potentially allowing a stellar-mass black hole explanation.

Finally, we obtain a maximum luminosity via an effective surface area.
We define the inner radius, $r_{0}$, of our line-emitting region by 
assuming the line-emitting gas is in a Keplerian orbit about central mass $M$ with velocity $v$ (which we relate to observed line-widths below) so that  
\begin{equation}
r_{0} =  GM/v^2.
\label{rnot}
\end{equation}
We assume the line-emitting region comprises an annulus with width comparable to $r_{0}$.
Line luminosities are written
\begin{equation}
L_{\lambda} = 4 \pi r_{0}^2 I_{\lambda} = 4 \pi \left(\frac{GM}{v^2}\right)^2 I_{\lambda}.
\label{eqn:lumline} 
\end{equation}
We evaluate constants, normalize variables to convenient values, and obtain maximum luminosity
\begin{equation}
L^{max}_{\lambda} =  L^{0}_{\lambda} {M_1}^2 {v_{100}}^{-4}.
\label{eqn:lumline2} 
\end{equation}
\noindent where $M_1$ is the central mass in units $\Msun$,
$v_{100}$ is the rotational velocity in units $100$~km~s$^{-1}$, and
all constants have been absorbed into fiducial values $L^{0}_{\lambda}$.
Table~\ref{tab:data} contains the fiducial luminosities for \oiii $\lambda5007$ and \nii $\lambda6584$. 
Atomic data used in this section are from Wiese, Fuhr, \&  Deters (1996), 
as obtained from the NIST Atomic Spectra Database (Ralchenko et al. 2008).

It is important to emphasize that equations~\ref{eqn:lumline} and \ref{eqn:lumline2} 
are independent of temperature, volume density, and metallicity.  
This is by design.
However, temperature and metallicity do inform the \emph{practicality} of the equations.
We will discuss the former below.  Regarding the latter, 
at fixed $n_e$, the disc height required to reach a certain column density (and therefore luminosity)
is inversely proportional to metallicity.  
A purely metallic plasma allows the smallest disc height for the maximum-luminosity configuration of our model.
Solar metallicity gas requires a disc height more than three orders of magnitude greater,
reaching $\approx10$ Thomson depths.
Scattering would yield emergent line widths $\sim1000$~km~s$^{-1}$.  
Such large widths would contradict our rotational broadening assumption in both ULXs
and are ruled out completely by observation in CXOJ033831.8-352604.
If the width scales as the number of scatterings (usually the greater of $\tau$ and $\tau^2$), 
this problem would be eliminated by decreasing the column by some factor.
The minimum central mass found below would increase by the square of that factor.

\begin{table}
\centering
\caption{Observed and model luminosities (erg~s$^{-1}$).  
The last column contains the $L^{0}_{\lambda}$ values defined via equation~\ref{eqn:lumline2}.}
\begin{tabular}{lrrr}
Quantity & CXOJ033831.8-352604 & RZ2109 & $L_{\lambda}^{0}$ \\
\hline
$L_{5007}$	& few $\times10^{36}$	& $1.4\times10^{37}$	& $2.4\times10^{33}$ \\  
$L_{6584}$	& $3\times10^{36}$		& unobs.				& $2.8\times10^{32}$ \\  
$L_x$ 	& $1.5-2.3\times10^{39}$		& $4\times10^{39}$		&  \\
\end{tabular}
\label{tab:data}
\end{table}

\begin{figure}
\protect\resizebox{\columnwidth}{!}
{\includegraphics[angle=0]{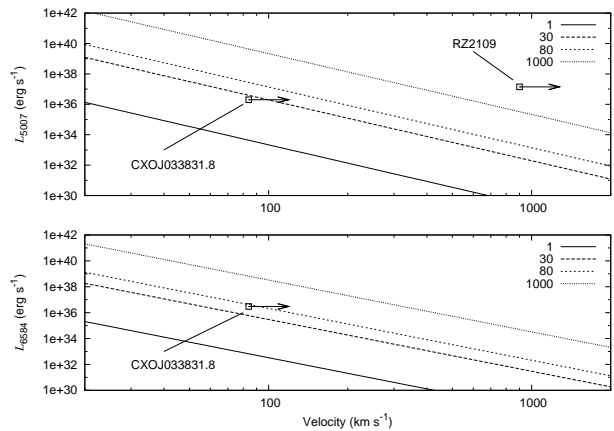}}
\caption{Maximum line luminosities (upper: \oiii $\lambda5007$; lower: \nii $\lambda6584$)
versus rotational velocity of the line-emitting region for several values
of the central mass (in units \Msun).
Squares indicate results from I10 and Z08.
Arrows point from minimum (edge-on) velocity to the velocity appropriate for $45\,^{\circ}$ inclination.  
\nii was not detected in RZ2109.}
\label{fig:maxL}
\end{figure}

We plot in Figure~\ref{fig:maxL} maximum line luminosities as a function of the velocity of the gas at $r_0$.
Several values of the compact object mass are considered.    
The middle two lines represent recent theoretical ($80 \Msun$) and observed ($30 \Msun$) upper limits to the mass of a stellar black hole (Belczynski  et al. 2010).  
Results for I10 and Z08 are indicated by the squares, 
where we have assumed velocities equal to $1/(2\sqrt{\ln2})$ times the measured FWHM.
This ignores the effect of disc inclination on the observed line widths.  
Accounting for inclination would mean a greater rotational velocity and a smaller annulus radius (for a given central mass).
This would decrease the surface area of the line-emitting region and the luminosity of the line.
The arrow on each square points to the velocity appropriate for the (arbitrarily chosen) inclination angle $45\,^{\circ}$.

The point corresponding to CXOJ033831.8-352604 in the upper panel of Figure~\ref{fig:maxL} is 
beneath the lines corresponding to the theoretical and observed stellar-mass black hole upper limits.
However, the lower panel of Figure~\ref{fig:maxL} makes a more compelling case.
The observed $L_{6584}$ is very slightly less than (at edge-on inclination) the maximum predicted luminosity 
corresponding to the maximum theoretical mass of a stellar black-hole (Belczynski  et al. 2010).
Note that these predictions assume optimal cooling efficiency.
I10 found $T_e \leq 13,000$~K in the \nii emitting region using the lack of $\lambda5755$.
Using this temperature would lower the \nii predictions by a factor $\sim2$ and require $M\gtrsim100 \Msun$.
\oiii $\lambda4363$ was neither presented nor mentioned in I10.  
Measurement  of this line could place a strong constraint on the temperature of the \oiii region.

The upper panel of Figure~\ref{fig:maxL} clearly demonstrates that the Z08 observations of RZ2109 
cannot be explained as Keplerian rotation around a stellar-mass object.  
The required mass is at least $6000 \Msun$.
\oiii $\lambda4363$ is weakly detected in RZ2109.
A rough estimate is that $\lambda5007$ is at least 10 times greater.
The Nussbaumer \& Storey (1981) tabulations would then give $T\lesssim15,000$~K at the $\lambda5007$ critical density.
That upper limit quickly decreases to $T\lesssim7,500$~K for $n_e \gtrsim10^7$~cm$^{-3}$.
These considerations depress the maximum luminosity by at least a factor of 20,
so that the minimum central mass is at least $30,000 \Msun$.

We rule out stellar black-hole Keplerian accretors in both sources.
Sub-Keplerian motions have orbiting radii that are smaller for a given central mass than Keplerian, 
decreasing the surface area of the line-emitting region and exacerbating the problems discussed above.
We can therefore rule out sub-Keplerian, stellar black hole accretion more readily than we can rule out Keplerian.

We stress that the two panels of Figure~\ref{fig:maxL} do not depend upon the relative abundances of oxygen and nitrogen.
The predicted \nii luminosities are noticeably less than the \oiii luminosities at the same velocity and central mass.
This is because the \nii lines are less efficient coolants than their \oiii counterparts.

\section{Total disc mass}
\label{discmass}

Here we present a bootstrap argument for the total disc masses in an accretion scenario.
The mass required to emit $L_{5007}$ (or $L_{6584}$) is 
\begin{equation}
M(^1D_2) = \frac{Z m_p L_{\lambda}}{h\nu A},
\end{equation}
\noindent where $Zm_p$ is the mass of a single atom with mass number $Z$.
This is independent of the geometrical model and the gas density (even below $n_{\mathrm{crit}}$),
and accounts only for the population in the upper level of the transition.
Observed values of $L_{\lambda}$ are listed in Table~\ref{tab:data}. 
The \oiii lines require masses $\sim3\times10^{-7}$ and $\sim2\times10^{-6}\Msun$ for CXOJ033831.8-352604 and RZ2109, respectively.
The \nii lines in CXOJ033831.8-352604 require $4\sim10^{-6}\Msun$.

\begin{figure}
\protect\resizebox{\columnwidth}{!}
{\includegraphics[angle=0]{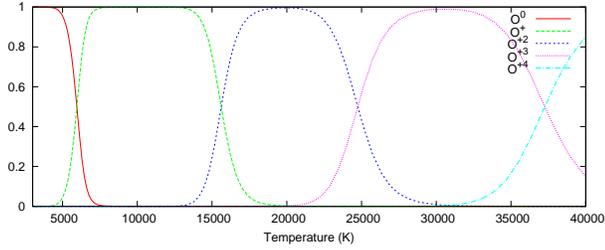}}
\caption{Ionization fractions of oxygen as a function of the temperature of an intense blackbody.}
\label{ionfracs}
\end{figure}

We can bootstrap our way to a total disc mass via 
\begin{equation}
\frac{M(^1D_2)}{M_{\mathrm{disc}}} = \frac{\rho_{\mathrm{Z}}}{\rho} \frac{\int f_{\mathrm{ion}} f(^1D_2) r dr}{\int r dr},
\label{bootstrap}
\end{equation}
where $f(^1D_2)$ was discussed in section~\ref{maxlum}
and $f_{\mathrm{ion}}$ is the C-like ionization fraction,
which we estimate using the well-known plasma simulation code Cloudy (version C08, last described by Ferland et al. 1998).
Figure~\ref{ionfracs} plots the results for oxygen
as a function of the temperature of a bright blackbody photoionization source.
The fraction in O$^{+2}$ peaks around $20,000$~K.
The maximum equivalent width of $\lambda5007$ with $N$(O$^{+2}$)$ = 10^{22}$~cm$^{-2}$ against a blackbody of the same area
and $T_{BB} = 20,000$~K is about $1$\AA.
We estimate from spectra in I10 and Z08 observed equivalent widths of $5$ and $15$\AA.
Blackbody sources with weak enough continuum at $\sim5000$\AA \ are too weak to ionize O$^{+}$.
We require a harder continuum.

\begin{figure}
\protect\resizebox{\columnwidth}{!}
{\includegraphics[angle=0]{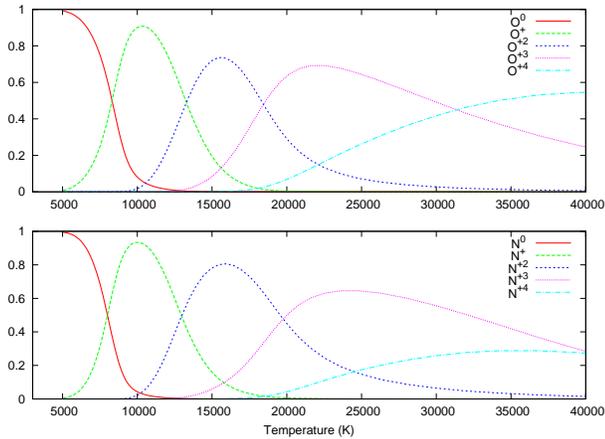}}
\caption{Ionization fractions of oxygen (upper panel) and nitrogen (lower panel) 
as a function of the electron temperature of gas ionized by a power-law continuum with spectral index $\Gamma = 2.5$.}
\label{ionfracs2}
\end{figure}

We instead consider, again using Cloudy, a power-law continuum with spectral index $\Gamma=2.5$ (following X-ray continuum fits in I10).
Figure~\ref{ionfracs2} plots both oxygen and nitrogen ionization fractions as a function of electron temperature.
The ionization parameter ranges from $-9 \leq \log U \leq 0$.

\begin{figure}
\protect\resizebox{\columnwidth}{!}
{\includegraphics[angle=0]{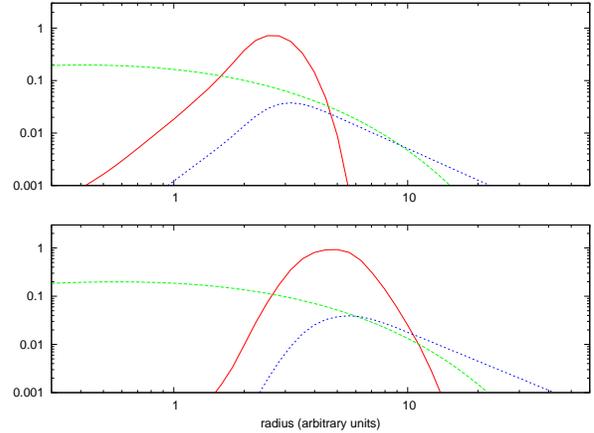}}
\caption{Level and ionization stage fractions of oxygen (upper panel) and nitrogen (lower panel) versus normalized radius.
Solid red lines are fractions in O$^{+2}$ and N$^{+}$, dashed green lines are $f(^1D_2)$,
and dotted blue lines are ratios of integrals given in the right-hand side of equation~\ref{bootstrap}.}
\label{accum}
\end{figure}

The fractions $f_{\mathrm{ion}}$ and $f(^1D_2)$ depend on temperature.
We relate them to the variable of integration $r$ via the standard Shakura-Sunyaev prescription ($T\propto r^{-3/4}$).
The ratio of integrals in equation~\ref{bootstrap} is illustrated by the dotted blue curves in Figure~\ref{accum}.
The result depends on the upper limit of integration.  
The peak is $\approx4\%$ for both oxygen and nitrogen.  
The curves decline as $r^{-2}$ at large radii.

An important conclusion can be derived from Figures~\ref{ionfracs2} and \ref{accum}.  
The regions emitting \oiii and \nii are mostly non-cospatial.  
The details depend on the ionizing source and the geometry.
Since the second ionization energy of nitrogen is only $85\%$ of the second ionization energy of oxygen,
the basic result is fairly robust.
N$^+$ always peaks at a temperature significantly less than the temperature of the O$^{+2}$ peak.
If the standard $\alpha$-disc is applicable, we have $T\propto r^{-3/4}$.
Keplerian rotation gives $v \propto r^{-1/2}$, so $v \propto T^{2/3}$.
A series of tests suggests the \oiii and \nii line widths should differ by a factor $\gtrsim1.4$.
The widths are uncertain but appear comparable in the I10 observations.
This argues against Keplerian rotation about \emph{any} central mass in CXOJ033831.8-352604.
A very steep temperature gradient or a thermal instability between the \oiii and \nii regions could mitigate this problem.
High-resolution spectroscopy would be useful in efforts to address such questions.

Finally, we require the density of oxygen relative to the total gas density.
Assuming metallicity is independent of radius, we can write
\begin{equation}
\frac{\rho_{\mathrm{Z}}}{\rho} = \frac{A~n_Z}{\Sigma~A~n_Z}
\end{equation}
where $A$ is atomic mass number.
For fixed relative metal abundances, 
the fraction of mass in oxygen varies from $\sim0.005$ with solar abundances (Grevesse \& Sauval 1998)
to $\sim0.4$ for all metals.

Combining the above results, we find $M_{\mathrm{disc}} \gtrsim10^{-4}\Msun$.
This corresponds to the purely metallic case.
If we assume roughly solar abundances, $M_{\mathrm{disc}} \gtrsim10^{-2}\Msun$.
An estimate for the upper limit is not readily apparent in either system
because we have no way of estimating the outer radius and, in particular,
because the line-emitting gas could be only a thin outermost layer of the disc.

\section{Roche-lobe overflow}
\label{roche}

In steady Roche-lobe overflow accretion,
the disc mass must be replenished on the viscous time-scale
\begin{equation}
t_{\mathrm{visc}} = \frac{M_{\mathrm{disc}}}{\dot{M}},
\label{tvisc}
\end{equation}
where the mass-accretion rate $\dot{M} = L_x / \eta c^2$ and
$\eta$ is an efficiency typically taken to be $0.1$.
We can also consider the viscous time-scale in a standard $\alpha$-disc.
We manipulate the familiar expression (Frank, King, \& Raine 2002), 
substituting azimuthal velocity for the radius via equation~\ref{rnot}, and obtain
\begin{equation}
t_{\mathrm{visc}} \approx 4.2 \alpha^{-4/5} {\dot{M}_{16}}^{-3/10} {M_1}^{3/2} {v_{100}}^{-5/2}~ \mathrm{yr},
\label{tvisc2}
\end{equation}
where ${\dot{M}_{16}}$ is the mass accretion rate in units $10^{16}$g~s$^{-1}$.
For the X-ray luminosities given in Table~\ref{tab:data}, 
we find $\dot{M}\approx2\times10^{19}$g~s$^{-1}$ ($3\times10^{-7}$\Msun~yr$^{-1}$).

If we assume the ``edge-on'' velocities discussed in relation to Figure~\ref{fig:maxL} and $\alpha=0.1$,
equations~\ref{tvisc} and \ref{tvisc2} require masses $70$ and $4000\Msun$
for CXOJ033831.8-352604 and RZ2109, respectively.
These numbers are surprisingly similar to the minimum masses found in Section~\ref{maxlum}.
The minimum masses are $\sim40$ times larger if the disc has $Z\approx0.3 Z_\odot$.
The viscous time-scale $t_{\mathrm{visc}}\gtrsim300$~yr.

Temperatures derived with the $\alpha$-disc prescription, standard assumptions, and the above limits
are too cold by a factor $\sim10$.
$M_1$, $\dot{M}$, and $\alpha$ all enter as fairly weak powers and offer no obvious solution.
Irradiation of a surface layer (e.g., Dubus et al. 1999) might yield the necessary temperatures.

Can a globular cluster form and sustain the massive accretion discs implied by the observed \oiii luminosities?
Combining the minimum masses obtained in Section~\ref{maxlum} with the rotational velocities derived above, 
we obtain $r_{0} \approx 2\times10^{14}$ and $5\times10^{14}$~cm,
for CXOJ033831.8-352604 and RZ2109, respectively.
We take this as the distance from the center of the compact object to the Lagrange point $L_1$.  
We then assume a $1 \Msun$ donor star (giving mass ratios $q = 0.01$ and $1.7\times10^{-4}$) 
and employ approximations from Eggleton (1983) and Frank, King \& Raine (2002),
and we obtain donor Roche lobe radii 
$r_L \gtrsim10^{13}$ and $5\times10^{12}$~cm, respectively.
These radii are comfortably within the size limits of the red giants that are common in globular clusters,
but they represent lower limits.
Larger central object masses would correspond to larger overflow radii
(because the dependence of $r_L/r_0$ on $M$ via $q$ is not enough to overcome $r_0\propto M$).
White dwarfs have been mentioned because of the apparent high-metallicity in both systems.
With prototypical $M_{wd} = 0.6\Msun$ and $r_{wd} = 10^9$~cm, 
we can \emph{strongly} exclude white dwarfs as Roche-lobe overflow donors in either system.

Does the envelope of a red giant contain enough oxygen to produce the observed line luminosities?
Carretta \& Gratton (1997) considered the metallicities of red giants in 24 galactic globular clusters.
They found sub-solar metallicities for every star in their study.  
The majority were sub-solar by at least an order of magnitude.
The most metal-rich had $Z\approx0.3 Z_\odot$.

The [OIII]/H$\beta$ and [NII]/H$\alpha$ values I10 observed in CXOJ033831.8-352604 ($\gtrsim5$ and $\gtrsim7$, respectively)
and the [OIII]/H$\beta$ value Z08 observed in RZ2109 ($\sim30$) are not necessarily indicative of high-metallicity gas.
The photoionized gas near the center of the giant HII region 30 Doradus has $Z\approx0.3 Z_\odot$ and [OIII]/H$\beta\approx6$ (Pellegrini et al., submitted).
Elevated ratios are often thought to be signs of radiative shock heating (e.g., Dopita \& Sutherland 1995),
as was noted by Z08.
Detailed attempts to distinguish between enhanced abundances and non-equilibrium heating processes 
are probably unwarranted without better observational constraints.
A further complication pertains to whether the line-emitting gas is radiation- or matter-bound (McCall, Rybski, \& Shields 1985),
although the high observed [NII]/H$\alpha$ would seem to make the latter less likely in CXOJ033831.8-352604.  

If Z08 are correct in their black-hole wind explanation of RZ2109,
we offer without comment the suggestion that 
the observed very broad line-widths may be the low-resolution appearance of 
narrow emission lines atop broader ones as in a two-wind structure (e.g., Fernandes 1999).

\section{Acknowledgments}

RLP thanks referee John Raymond for excellent suggestions and criticisms and Joel Bregman, Gary Ferland, Jimmy Irwin, Jon Miller, Eric Pellegrini,
Mark Reynolds, and Pete Storey for fruitful discussions.

\bsp

\label{lastpage}
\clearpage

\begin{thebibliography}{99}
\label{bibliography}

\bibitem[\protect\citeauthoryear{Andre\"{a}, Dreschel, \& Starrfield}{1994}]{ADS94}
Andre\"{a} J., Dreschel H., Starrfield S., 1994, A\&A,  291, 869

\bibitem[\protect\citeauthoryear{Belczynski  et al.}{2010}]{Belczynski10}
Belczynski K., Bulik T., Fryer C. L., Ruiter A., Valsecchi F., Vink J. S., Hurley J. R. 2010, ApJ, 714, 1217

\bibitem[\protect\citeauthoryear{Carretta \& Gratton}{1997}]{CG97}
Carretta E., Gratton R. G. 1997, A\&AS, 121, 95

\bibitem[\protect\citeauthoryear{Dopita \& Sutherland}{1995}]{DS95}
Dopita M. A., Sutherland R. S., 1995, ApJ, 455, 468

\bibitem[\protect\citeauthoryear{Dubus et al.}{1999}]{D99}
Dubus G., Lasota J.-P., Hameury J.-M., Charles P., 1999, MNRAS,  303, 139

\bibitem[\protect\citeauthoryear{Eggleton}{1983}]{Eggleton83}
Eggleton P., 1983, ApJ, 268, 368

\bibitem[\protect\citeauthoryear{Fabian}{1975}]{F75}
Fabian A. C., Pringle J. E., Rees M. J. 1975, MNRAS, 172, 15

\bibitem[\protect\citeauthoryear{Ferland et al.}{1998}]{Ferland98}
Ferland G. J., Korista K. T., Verner D. A., Ferguson J. W., Kingdon J. B., Verner E. M. 1998, PASP, 110, 761

\bibitem[\protect\citeauthoryear{Fernandes}{1999}]{Fernandes99}
Fernandes R. C., Jr. 1999, MNRAS, 305, 602

\bibitem[\protect\citeauthoryear{Frank, King, \& Raine}{2002}]{FKR02}
Frank J., King A., Raine D., 2002, 
Accretion Power in Astrophysics, 3rd.~ed.~Cambridge, UK: Cambridge University Press

\bibitem[\protect\citeauthoryear{Grevesse \& Sauval}{1998}]{GS98}
Grevesse N., Sauval A., 1998, Space Science Reviews, 85, 161

\bibitem[\protect\citeauthoryear{Hui \& Krolik}{2008}]{HK08}
Hui Y., Krolik J., 2008, ApJ, 679, 1405

\bibitem[\protect\citeauthoryear{Jose \& Hernanz}{1998}]{JH98}
Jos\'{e} J., Hernanz M., 1998, ApJ,  494, 680

\bibitem[\protect\citeauthoryear{Kastner \& Bhatia}{1989}]{KB89}
Kastner S. O., Bhatia A. K., 1989, ApJSS, 71, 665

\bibitem[\protect\citeauthoryear{Keenan \& Aggarwal}{1987}]{KA87}
Keenan F. P., Aggarwal, K. M. 1987, ApJ, 319, 403 

\bibitem[\protect\citeauthoryear{King}{2008}]{K08}
King A. R., 2008, MNRAS, 385, 113

\bibitem[\protect\citeauthoryear{King}{2009}]{K09}
King A. R., 2009, MNRAS, 393, 41

\bibitem[\protect\citeauthoryear{Irwin et al.}{2010}]{I10}
Irwin J. A., Brink T. G., Bregman J. N., Roberts T. P., 2010, ApJ, 712, 1

\bibitem[\protect\citeauthoryear{McCall, Rybski, \& Shields}{1985}]{MRS85}
McCall M. L., Rybski P. M., Shields G. A., 1985, ApJS, 57, 1

\bibitem[\protect\citeauthoryear{Miller, Fabian, \& Miller}{2004}]{MFM04}
Miller J. M., Fabian A. C., Miller M. C., 2004, ApJ, 614, 117

\bibitem[\protect\citeauthoryear{Nussbaumer \& Storey}{1981}]{NS81}
Nussbaumer H., Storey P. J., 1981, A\&A, 99, 177

\bibitem[\protect\citeauthoryear{Osterbrock \& Ferland}{2006}]{AGN3}
Osterbrock D. E., Ferland G. J., 2006,
Astrophysics of gaseous nebulae and active galactic nuclei, 2nd.~ed
.~by D.E.~Osterbrock and G.J.~Ferland.~Sausalito, CA: University Science Books

\bibitem[\protect\citeauthoryear{Pellegrini et al.}{2010}]{P10}
Pellegrini E. et al., submitted

\bibitem[\protect\citeauthoryear{Ralchenko et al.}{2008}]{Ralchenko08}
Ralchenko Yu., Kramida A. E., Reader J., NIST ASD Team (2008). NIST Atomic Spectra Database (version 3.1.5), [Online]. Available: http://physics.nist.gov/asd3  [2010, March 29]. National Institute of Standards and Technology, Gaithersburg, MD. 

\bibitem[\protect\citeauthoryear{Rosswog et al.}{2009}]{Rosswog09}
Rosswog S., Ramirez-Ruiz E., Hix W. R. 2009, ApJ, 695, 404

\bibitem[\protect\citeauthoryear{Rybicki \& Lightman}{1979}]{RL79}
Rybicki G. B., Lightman A., 1979, Radiative Processes in Astrophysics, New York, Wiley.

\bibitem[\protect\citeauthoryear{Shakura \& Sunyaev}{1973}]{SS73}
Shakura N. I., Sunyaev R. A., 1973, A\&A, 24, 337

\bibitem[\protect\citeauthoryear{Wiese, Fuhr, \&  Deters}{1996}]{Wiese96}
Wiese W. L., Fuhr J. R., Deters T. M., 1996, J. Phys. Chem. Ref. Data, Monograph No. 7

\bibitem[\protect\citeauthoryear{Winter, Mushotzky, \& Reynolds}{2006}]{WMR06}
Winter L. M., Mushotzky R. F., Reynolds C. S., 2006, 649, 730

\bibitem[\protect\citeauthoryear{Zepf et al.}{2007}]{Z07}
Zepf S. E., Maccarone T. J., Bergond G., Kundu A., Rhode K. L., Salzer J. J., 2007, ApJ, 669, 69

\bibitem[\protect\citeauthoryear{Zepf et al.}{2008}]{Z08}
Zepf S. E., Stern D., Maccarone T. J. et al., 2008, ApJ, 683, 139

\end{thebibliography}
\end{document}